\documentclass[preprint,aps,floatfix,showpacs]{revtex4}
\usepackage{graphics, epsfig ,color} 
\usepackage{amsmath,amssymb} 

\begin{document}
 \date{\today}
 \title{Cavity Soliton Laser based on mutually coupled semiconductor microresonators}
 \author{P. Genevet, S. Barland, M. Giudici and J.R. Tredicce }
 \affiliation{Universit\'e de Nice Sophia Antipolis, Institut Non-Lin\'eaire de Nice, UMR 6618,06560 Valbonne, France}

\begin{abstract}
We report on experimental observation of localized structures in two mutually
coupled broad-area semiconductor resonators. These structures coexist with a dark
homogeneous background and they have the same properties as cavity solitons without
requiring the presence of a driving beam into the system. They can be switched
individually on and off by means of a local addressing beam.
\end{abstract} 


\maketitle

\par
Localized Structures (LS) form in large aspect-ratio media where two or several
solutions coexist in the parameter space (see \textit{e.g.} \cite{akhmediev} for a recent review). Cavity Solitons (CS) are
LS generated in a cavity filled with a non linear medium driven by a coherent
injected field (holding beam, HB) where they appear as single bistable bright intensity
peaks coexisting with a homogeneous background. Their existence and mutual independence in semiconductors have been
experimentally demonstrated in microcavities operated as optical amplifiers
\cite{solitette,lugiatoreview2003} since a local perturbation in
form of a beam coherent with the HB can be used for switching CS on and off
independently \cite{solitette,hachair2004}. The possibility to control their
location and their motion by introducing phase or amplitude gradients in the
holding beam suggests their use as mobile pixels for all-optical processing
units. Indeed, in the last decade, CS in semiconductor have attracted a growing
interest since they combine the bistability and plasticity properties with the
advantages of semiconductor media in terms of fast response and small size.
 The application potential of CS has been evidenced with some
first-principle demonstration of new all-optical devices exploiting CS
properties for optical memories \cite{positioning06} and delay lines
\cite{ApplPhysLett92_011101}.  Nevertheless, the tight conditions required for
CS stability in present experimental schemes hinders their application to non
prototypical devices. A radical simplification could be achieved implementing
the concept of Cavity Soliton Laser, \textit{i.e.} a device generating CS without an
external injection beam.  Some steps in this direction have been made recently
with a scheme based on broad area VCSEL submitted to frequency selective
feedback \cite{tanguy:013907}. However, even if no HB is present it appears that the
stability of localized structures in this case depends critically on feedback
alignment and on the detuning between the resonator and the external frequency selecting element. An alternative
approach is provided by a laser with a saturable absorber (SA). Indeed, this
system is among the first ones theoretically shown to possess the necessary
ingredients for the generation of localized structures in optics, called in
this case dissipative autosolitons \cite{rosanov1992}. While this initial work was
realized in the limit of fast materials
\cite{rosanov1992,rosanov_spatial_hysteresis}, it was later extended to the
case of finite relaxation times \cite{PhysRevE.61.5814,rosanov_spatial_hysteresis}.
Finally, the case of slow absorber material (as would be the case for
semiconductors) was examined in \cite{bache2005} and the case where the
absorbing and gain media have equal response times has been  studied in
\cite{prati2007}.  The authors of these last references show numerically that
in a vertical cavity surface emitting laser (VCSEL) with saturable absorber, CS
related to the existence of a modulational instability can be switched on and
off by injecting a local optical perturbation.

Despite this extensive theoretical and numerical research, we are not aware of
any experimental observation of CS using a saturable absorber in semiconductor
laser.  In this letter we show the experimental realization of a CSL based on
two mutually coupled micro resonators where one plays the role of an amplifier
and the second of a saturable absorber. As we shall demonstrate below, this
scheme allows a remarkably  simple realization of a cavity soliton laser. We
show that  bistable solitary structures coexist with a dark
homogeneous background and that they can be switched on and off independently by an
incoherent beam. We believe that this demonstration opens the way towards
implementations in very compact monolithic devices including both the amplifying and saturable absorber sections \cite{fischer:3020}.

\par

\begin{figure}[h!]
 {\includegraphics[width=0.45\textwidth]{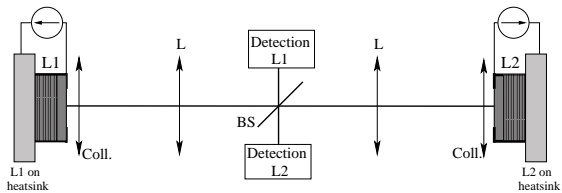}}
\caption{Schematic drawing of the experiment. $L_1$: Laser above the transparency, $L_2$: Laser
below the transparency, BS: beam splitter. Detection of $L_1$ (resp. $L_2$) includes a CCD camera monitoring the near field of $L_1$ (resp. $L_2$) and a
fast detector to monitor the local temporal behavior.} \label{schema}
\end{figure}

The lasers we use are two nominally identical VCSELs provided by ULM Photonics. They are
oxydized bottom-emitter VCSELs  emitting around $975 nm$ \cite{grabherr98}. Their
transverse section is 200~$\mu$m. They are mounted in a mutually coupled
configuration, where one laser is electrically biased above the transparency but
below its standalone coherent emission threshold ($L_1$), while the other is
biased below transparency ($L_2$). Both devices packages are temperature
stabilized.  In order to compensate the effects of diffraction during the
propagation in the extended cavity (60~cm long) and to keep the
system highly symmetric, we use identical collimators and identical lenses
placed such that the two resonators are in a self imaging condition (see Fig.~
\ref{schema}). This configuration allows to preserve the high Fresnel number
required for the existence of LS \cite{lugiatoreview2003}. A $20~\%$ reflection
beam splitter is inserted in the center of the cavity to extract two output
beams from the system. Time-averaged near-field profiles of both resonators are
simultaneously imaged on two charge-coupled device (CCD) cameras. The use of
two CCD cameras allows to check the output of each element of the coupled system independently. In the detection path of laser $L_1$, an iris placed in an intermediate near field plane enables to  select a small area of the profile for
point-like temporally resolved detection. A photodetector Thorlabs PDA8GS (less
than 100~ps rise time) coupled to a digital oscilloscope LeCroy Wavemaster
8600A (6~GHz analog bandwidth) monitors the output of this small portion of the
$L_1$ profile.

\par

\begin{figure}[h!]
{\includegraphics[width=0.4\textwidth]{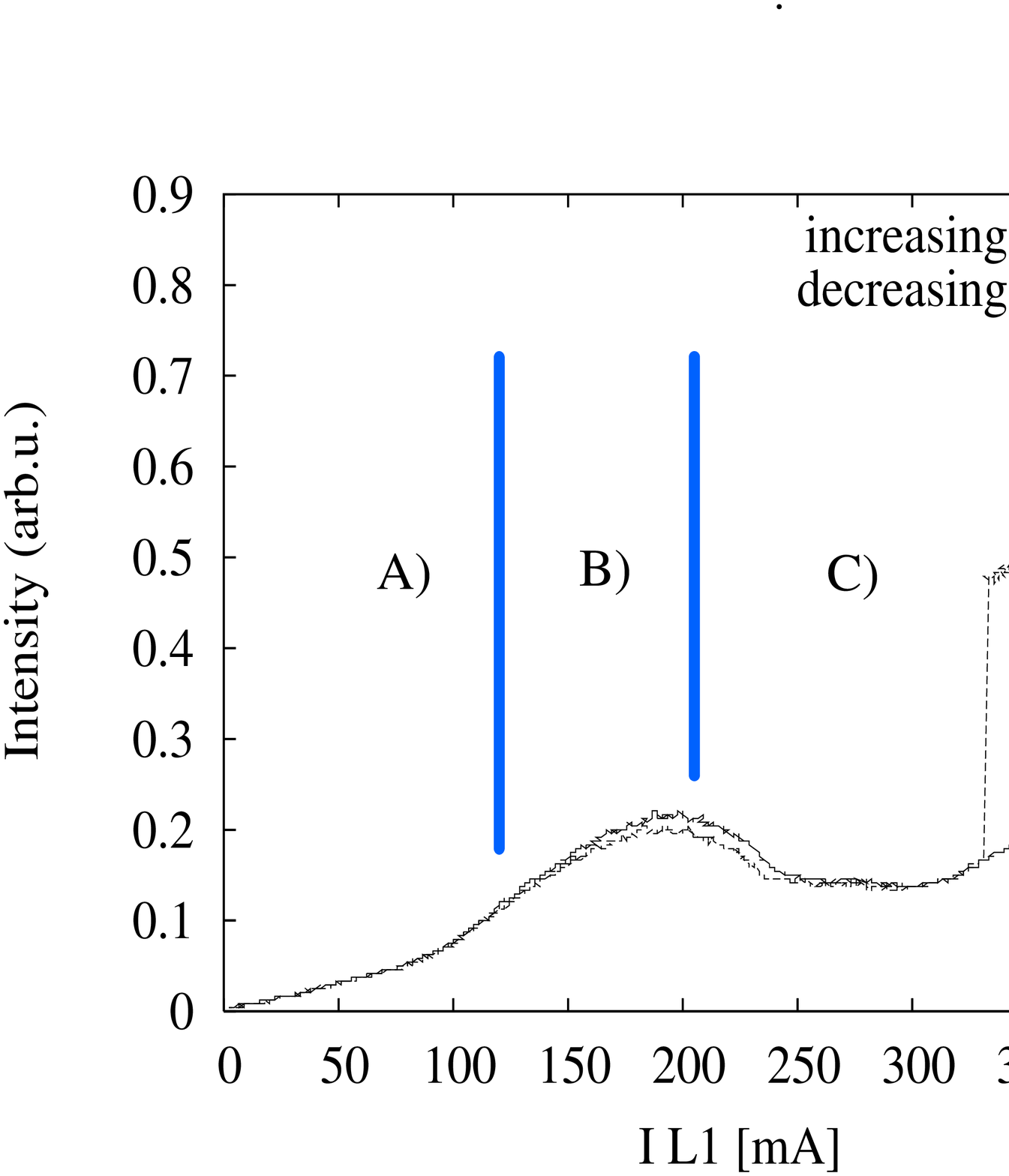}}
\caption{Local intensity output emit by the system when we scan $I_{L_1}$ for  all
the other parameters constant ($I_{L_1}$ is drive very slowly to keep the
temperature controller stable). A) below threshold, B)lasing by
feedback, C)Absorption by $L_2$, bistable behavior, D) pattern formation.}
\label{bistability}
\end{figure}

The solitary VCSELs light intensity output as function of the pumping current
($I_{L_1,L_2}$) indicate that the uncoupled lasers have very similar
standalone coherent emission thresholds $I_{L_1}^{th} \sim I_{L_2}^{th} \sim 400
mA $. The temperature control of each device is set such that the emission
wavelength of laser $L_1$ is approximately 1~nm blue detuned  with
respect to laser $L_2$ when both devices are pumped by the same amount of
current. In Fig.~\ref{bistability} we show the local intensity output of the
compound system as a function of pumping current of $L_1$, ($I_{L_1}$), while
$I_{L_2}$ is kept fixed at a few mA, \textit{i.e.} below the transparency
pumping value. The monitored region has a diameter of about 20~$\mu m$ and it is
placed in the centre of the device. When $I_{L_1}\leq 100~mA$ (zone A in
Fig.~\ref{bistability}, the increase of emitted power is attributed to
spontaneous emission, since the optical spectrum of the system does not show
coherent emission.  The first threshold ($I_{th}\sim 100~
mA$, region~B of Fig.~\ref{bistability}) is reached when light linearly reflected
on the output mirror of $L_2$ sufficiently reduces the losses of the compound
system such that laser emission can be obtained. At that point, threshold reduction and coherent
emission may therefore be attributed to losses reduction as is shown in
conventional optical feedback experiments.  

For, $I_{L_1} \geq 200 mA$ (region~C of Fig.~\ref{bistability}) the power output
of the compound system saturates and then decreases for increasing values of
$I_{L_1}$. This happens when the longitudinal resonances of both cavities match
and absorption in device $L_2$ takes place. The fact that the resonance
frequencies of both devices match only for certain current values is due to the
linear shift experienced by the laser frequency as a function of the pumping
current due to Joule heating.  The
absorption by laser $L_2$ in region~C of Fig.~\ref{bistability} has been
verified by performing the same measurement with laser $L_2$ unpumped, which
allows to verify the presence of a light induced current through the device,
which is absent in regions~A or~B. Increasing further $I_{L_1}$, the intensity
output remains constant at a low value. The optical spectra of this low
intensity emission shows a broad band peak indicating that this state
corresponds to spontaneous emission since absorption cut the feedback from $L_2$.  Further increase of $I_{L_1}$ above a critical value
$I_{L_1}^c$ the local intensity jumps up to a high value. The optical spectrum associated to this state shows a well pronounced
peak red-detuned with respect the spontaneous emission peak observed in the low
level state. In the near field profile this local transition of the intensity
output corresponds to the formation of a bright single peak structure inside the
monitored area. For larger $I_{L_1}$ (region~D of Fig.~
\ref{bistability}) the local intensity keeps increasing, while
the near field profile reveals the formation of multi peaked structures and
extended patterns around the monitored region. If $I_{L_1}$ is decreased the
local intensity shows hysteresis demonstrating bistability between a low and a
high emitted intensity state in the region monitored by the detector.


We note that while the curve shown in Fig.~\ref{bistability} was obtained for a
particular setting of the temperature of each device's substrate, equivalent
results may be obtained for different temperature and current settings provided
a number of conditions are satisfied. First of all, if device $L_2$ is pumped
at a too high value (that we interpret as above transparency??), we made no
observation of the bistable response of the system. Second, if $L_1$ is
initially red detuned with respect to $L_2$, these results won't be observed
since both devices won't be resonant for any current value of $L_1$. The
essential conditions are therefore an initial detuning allowing for current
induced tuning of $L_1$ towards the resonance of $L_2$ when $L_2$ is kept in an
absorbing regime. An additional condition is that if the suitable tuning
condition is obtained for a too low value of pumping of device $L_1$ (pumping of
device $L_2$ being kept constant) the total amount of
field in the compound system does not seem to reach the saturation value. In
this case, although the effects of absorption in $L_2$ can clearly be
identified, no bistable behavior was observed. We expect however that this
condition, related to the saturation fluence of device $L_2$, could also be
satisfied by applying a demagnification factor in the imaging of device $L_1$ on device $L_2$ in our setup or decreasing the reflectivity of the central beam splitter.

A detailed study of required conditions has been performed and will be
presented elsewhere, but for now we underline that provided that the previous
conditions are fulfilled, the curve shown on Fig.~\ref{bistability} can be
obtained for different settings of device temperatures and currents, with $I_{L_2}$ allowing to tune the amount of absorption in the system. As an
order of magnitude, the results presented here could be obtained for a broad
range of temperature settings leading to the bistability cycle being observed
for currents in $L_2$ ranging from 5 to 30~mA and currents in $L_1$ ranging
from 120 to 380~mA.

\begin{figure}[h!]
\centering
\includegraphics[width=0.4\textwidth,angle=0]{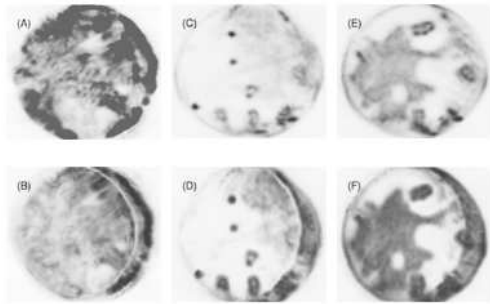}
\caption{Examples of near field of both devices. Dark areas correspond to high
intensities. a)Near field of the high current laser ($L_1$) before the
interaction, b) Near field of the low current laser ($L_2$) before the
interaction ($I_{L_1}=180~mA$), c,d) Near field of $L_1$ NF and $L_2$ in the
absorption zone ($I_{L_1}=358~mA$), e,f) Near field of $L_1$ and $L_2$ when the
pattern is developed ($I_{L_1}=365~mA$). $L_2$ is slightly shifted on the left. }
\label{Pattern}
\end{figure}

In Fig.~\ref{Pattern} we show the near-field transverse profiles of $L_1$ and
$L_2$ for different values of $I_{L_1}$, while $I_{L_2}$ is kept constant at 15~mA.
As discussed previously, in the self imaging scheme configuration, $L_1$ and $L_2$ are placed on self-conjugate
planes with a magnification of one. In order to monitor the absorption, the
devices are slightly shifted with respect to each other in the horizontal
direction. This way, a small portion close to the border of $L_1$ will not
interact with the corresponding portion of $L_2$ and it will be simply reflected
back by the substrate of the device. When $I_{L_1}$ is below the compound system's
threshold both profiles (not shown) are homogeneous. When $I_{L_1}$ is
increased above the first threshold (corresponding to region~B in Fig.~
\ref{bistability}, we observe the formation of complex and in general nonstationary patterns
resulting from the linear feedback effect of $L_2$'s output mirror on $L_1$ (Fig.~\ref{Pattern}
a),b)). In Fig.~\ref{Pattern}, b) showing the near field profile of $L_2$, the
dark area in the rightmost part of the image corresponds to high intensity
emitted by $L_1$ and reflected by the substrate of the device $L_2$ due to the lateral shift
we introduced. The bright arc nearby, closer to the center, corresponds to the edge of
$L_2$.

Keeping increasing $I_{L_1}$ the two resonators start to interact with $L_2$
absorbing the field emitted by $L_1$ (region~C of Fig.~\ref{bistability}) and the
near field profile of both lasers is mostly dark and homogeneous  except for
the small portion close to the rightmost edge, due to reflection onto the
substrate of $L_2$. By increasing the current
$I_{L_1}$ above $I_{L_1}^c$, bistable bright spots appear
spontaneously (see Fig.~\ref{Pattern} C,~D). 


Further increase of $I_{L_1}$ leads to the formation of generally nonstationary
filaments connecting the isolated spots together. For higher values of $I_{L_1}$
(Fig.~\ref{Pattern}~E and~F), corresponding to region~D of
Fig.~\ref{bistability}), a complex pattern develops progressively through the
whole transverse section.
Temporally resolved detection reveals that the widely spread pattern is generally not
stationary and exhibits  complex dynamics.

\begin{figure}[h]
\centering
\includegraphics[width=0.4\textwidth]{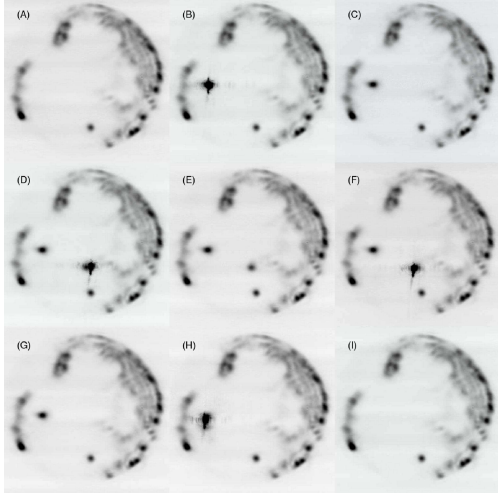}
\caption{$L_1$ near field intensity distribution. Dark areas correspond to high
intensities. Sequence of successive switching of two independent structures with an
incoherent WB tuned at the CS wavelength with all parameter fixed. a) Both
structures are off, b) The injection of WB switches one structure, c) the structure
remains on after the WB is blocked, d) the injection of the WB at a second location
ignites a second structure, e) when the WB is blocked, both structures are on, f)
reapplying the WB in the vicinity of the second structure attracts it to a slightly
different location, g) when the WB is blocked the structure switches off, h)  the WB
is applied in the vicinity of the first structure to attract it to a slightly
different location, i) when the WB is blocked, both structures are off.}
\label{independantCS}
\end{figure}

When the parameters are set in region~C of figure \ref{bistability}, the
observed bright isolated spots (Fig.~\ref{Pattern}, C,~D) are candidates for an
interpretation in terms of CS, since they coexist with a homogeneous background as shown by the hysteretical behavior as a function of $I_{L_1}$ (Fig.\ref{bistability}. While they appear
rather uncorrelated one to the other on the near field images, the full
demonstration of their mutual independence can only be performed by switching
them on and off with a local perturbation. Indeed, it has been shown
numerically that localized states in a cavity soliton laser can be switched on
or off by a coherent  \cite{bache2005,prati2007} or incoherent
\cite{aghdami2008} local optical perturbation. In our experiment, we use a
coherent beam whose optical frequency is close (here, within 0.1~nm) to the
emission frequency of the coupled device and whose diameter is about 15 $\mu m$. This beam is obtained from  an external-cavity tunable laser in Littman
configuration and applied on device $L_2$. By means of this local optical injection,
we are able to demonstrate independent switching of localized structures as
shown on Fig.~\ref{independantCS}.

The system is prepared in the low level emission state with the parameters set
in the bistable region. Starting with no spot, applying the WB to a point in
the transverse plane of $L_2$, we generate a high intensity spot with a
diameter of the order of $10 \mu m$. We then remove the WB and the bright spot
remains on indefinitely. We then apply the WB at a different position and a
second spot is generated without disturbing the first one, provided the
distance between the two spots is sufficient (no distance smaller than 40 $\mu m$ was observed). The two spots stay on even after
the WB is removed. We note that, as in previous experiments in semiconductor
devices \cite{hachair2004,tanguy:013907}, local device inhomogeneities seem to
play a role in the stabilization of localized states since although it is
possible to observe stable structures in different positions, not all positions
appear to be stable. We make use of this to switch off the localized
structures. In experiments involving an external forcing by a holding beam, the
simplest procedure is to apply a coherent local perturbation with opposite
phase with respect to the holding beam. In the present case, this approach is
not possible. Therefore we take advantage of mobility properties of the
structures to switch them off: by applying the same perturbation as before
close to a CS, we can drag it to a region of space where it is unstable and
therefore switches off (Fig.~\ref{independantCS}, F,~H). It is also possible to
switch off the structures without dragging them outside of their preferred
location if the pumping current of each device is set such that the system is
very close to the lower edge of the bistability region for the structure under
consideration. In this case though, our optical perturbation did not appear to
be sufficient to switch on a localized structure. Conversely, if the system
parameter were set very close to the upper edge of the bistability region, it
was not possible to switch localized structures off, except with the dragging
procedure described above. While the application of the local perturbation was
quasi continuous in the measurements shown above, numerical simulations
performed on a model for a monolithic semiconductor laser with saturable
absorber \cite{aghdami2008} indicate the possibility of ns switching time.
Preliminary experimental observations indicate that perturbations as short as
100~ns (limited by the modulator) can be sufficient to switch on or off a
localized structure, although no optimization (\textit{e.g} tuning and power of
the writing beam) has been performed yet.

In conclusion, we have given evidence of single localised peaks that fulfill all the
criteria required to be interpreted as cavity solitons in a compound semiconductor laser system with a saturable absorber. Since no injection beam is present in our
experiment we believe it is a very promising realization of a semiconductor cavity soliton laser. One of the greatest strengths of this experiment is its possibly straightforward miniaturization to monolithic devices able to generate self localized, bistable and mobile laser beams. In the present version of the system, the observation of spatially localized periodically pulsed transient regimes suggests the suitability of the system to the generation of three dimensional localized structures.

This work was supported by the FET Open Project  FunFACS (www.funfacs.org). We
are grateful to L.~Gil,  L.~Columbo  and G.~Tissoni for many useful
discussions.


\end{document}